\documentclass[11pt,a4paper,english,showpacs,superscriptaddress,aps,prd,nofootinbib]{revtex4-1}
\usepackage{amssymb}
\usepackage{amsmath}
\usepackage{graphicx}
\usepackage{babel}
\usepackage{hyperref}
\usepackage{color}
\usepackage{cancel}
\usepackage[normalem]{ulem}

\begin{document}

\title{Soft supersymmetry breaking in the nonlinear sigma model}

\author{L.~Ibiapina~Bevilaqua}
\email{leandro@ect.ufrn.br}
\affiliation{Escola de Ci\^encias e Tecnologia, Universidade Federal do Rio Grande do Norte\\
Caixa Postal 1524, 59072-970, Natal, Rio Grande do Norte, Brazil}

\author{A.~C.~Lehum}
\email{lehum@ufpa.br}
\affiliation{Faculdade de F\'isica, Universidade Federal do Par\'a, 66075-110, Bel\'em, Par\'a, Brazil.}

\author{A.~J.~da~Silva}\email{ajsilva@fma.if.usp.br}\affiliation{Instituto de F\'{\i}sica, Universidade de S\~{a}o Paulo\\ Caixa Postal 66318, 05315-970, S\~{a}o Paulo, S\~{a}o Paulo, Brazil}


\begin{abstract}

In this work we discuss the dynamical generation of mass in a deformed ${\cal N}=1$ supersymmetric nonlinear sigma model in a two-dimensional ($D=1+1$) space-time. We introduce the deformation by imposing a constraint that softly breaks supersymmetry. Through the tadpole method, we compute the effective potential at leading order in $1/N$ expansion showing that the model exhibit a dynamical generation of mass to the matter fields. Supersymmetry is recovered in the limit of the deformation parameter going to zero.   

\end{abstract}

\maketitle

\section{Introduction}

While the Nonlinear Sigma model (NLSM) has applications as a theory for the interaction between pions and nucleons~\cite{GellMann:1960np} and, in lower dimensional systems, it can also describe several aspects of condensed matter physics (for example, applications to ferromagnets~\cite{Polyakov:1975rr,Brezin:1975sq,Brezin:1976qa,Bardeen:1976zh,Sierra:1996nu}), the model is also appealing for purely theoretical investigations. In particular, it possesses an interesting phase structure and at the same time it shares some special features with more realistic theories, being a simple example of an asymptotically free theory~\cite{Friedan:1980jf,Hikami:1980hi}. 

The action for the NLSM in $D$ space-time dimensions may be written as
\begin{equation}\label{eq0a} 
S=\int\!{d^Dx}~\Big{\{}\frac{1}{2}\phi_a \Box\phi_a +\sigma\left(\phi_a^2-\frac{N}{g}\right)\Big{\}}~, 
\end{equation}
where the field $\sigma$ is a Lagrange multiplier that constraints the fields $\phi_a$ to satisfy $\phi_a^2=\dfrac{N}{g}$, such that the model has an $O(N)$ symmetry (the index $a$ assumes the values $1,2,...,N$). 

The phase structure and the renomalizability of the NLSM in $(2+1)$ dimensions was established by the late 1970s, showing that this model possesses two phases~\cite{Arefeva:1979bd,Arefeva:1978fj}. One phase is $O(N)$ symmetric and exhibits a spontaneous generation of mass due to a non-vanishing vacuum expectation value (VEV) of the Lagrange multiplier field $\sigma$, i.e., $\langle\sigma\rangle\ne0$. The other phase is characterized by a non-vanishing VEV of the fundamental bosonic field $\phi$, so that the $O(N)$ symmetry is spontaneously broken to $O(N-1)$, and there's no generation of mass. Several extensions of this model was later studied showing no changing in its phase structure~\cite{Arefeva:1980ms,Rosenstein:1989sg,Koures:1990hc,Koures:1991zu,Girotti:2001gs,Girotti:2001ku,Matsuda:1996vq,Jack:2001cd,Nitta:2003dv}. Unlike the two-phase structure of the 3D model, in two dimensions we have supersymmetry and $O(N)$ symmetry both unbroken, in agreement with a theorem by Coleman that states that in two dimensions Goldstone's theorem does not end with two alternatives (either manifest symmetry or Goldstone boson) but with only one: manifest symmetry \cite{Coleman}.

Although the supersymmetric counterpart of (\ref{eq0a}) presents a similar phase structure in $(2+1)$ dimensions, it was pointed out in \cite{Lehum:2013rpa} that there's no soft transition to the bosonic model for the mass acquired by the fields in the symmetric phase. To understand their point, consider the ${\cal N}=1$ SUSY NLSM, described by the action
\begin{eqnarray}\label{eq1}
S=\int\!{d^4z}~\Big{\{}\frac{1}{2}\Phi_a(z) D^2\Phi_a(z) +\Sigma(z)\left[\Phi_a(z)^2-\frac{N}{g}\right]\Big{\}}~,
\end{eqnarray}
where $2 D^2 = D^\alpha D_\alpha$, $D_\alpha = \partial_\alpha + i \theta^\beta \partial_{\alpha\beta}$ is the covariant supersymmetric derivative\footnote{As it is well known, the superspace in three and two space-time dimensions have the same structure~\cite{Gomes:2011aa}. For our notation conventions, please see Ref. \cite{Gates:1983nr}} and $\Sigma$ is the Lagrange multiplier superfield that constraints $\Phi_a$ to satisfy $\Phi_a^2(z)=\dfrac{N}{g}$.

If we write the superfields components as:
\begin{eqnarray}\label{eq1a}
&&\Phi_a(x,\theta)=\phi_a(x)+\theta^{\beta}\psi_{a\beta}(x)-\theta^2~F_a(x)~;\nonumber\\
&&\Sigma(x,\theta)=\rho(x)+\theta^{\beta}\chi_{\beta}(x)-\theta^2~\sigma(x)~,
\end{eqnarray}
we can integrate over $d^2\theta$, and eliminate the auxiliary field $F_a$ using its equation of motion, to express the action of the model as
\begin{eqnarray}\label{eq1c}
S=\int\!{d^2x}\Big{\{}\frac{1}{2}\phi_a\Box \phi_a +\frac{1}{2}\psi^{\alpha}_{a}i{\partial_{\alpha}}^{\beta}\psi_{a\beta}
+\sigma\left(\phi_a^2-\frac{N}{g}\right)-2\rho^2\phi_a^2+\rho\psi_a^{\beta}\psi_{a\beta}
+2\chi^{\beta}\psi_{a\beta}\phi_a\Big{\}}~,
\end{eqnarray}
and see that the auxiliary field $\sigma$ acts as the Lagrange multiplier associated to the constraint $\phi_a^2=\dfrac{N}{g}$ so that the usual (bosonic) model (\ref{eq0a}) is obtained setting $\psi=\rho=\chi=0$, and $\sigma \neq 0$.

From (\ref{eq1c}) it is easy to see that if exist a phase where mass is generated to the fundamental fields $\phi$ and $\psi$, their masses will be given by the VEV of the fields $\rho$ and $\sigma$ as
\begin{eqnarray}\label{eq1d}
M^2_{\phi}=4\langle\rho\rangle^2-2\langle\sigma\rangle~,\hspace{1cm} M^2_{\psi} = 4\langle\rho\rangle^2~,
\end{eqnarray}
\noindent
from which we observe that, for $\langle\sigma\rangle=0$ and a non-vanishing VEV of $\rho$, the fundamental bosonic and fermionic fields acquire the same squared mass $4\langle\rho\rangle^2$, indicating generation of mass in a supersymmetric phase as is well-known~\cite{Koures:1990hc,Koures:1991zu,Girotti:2001gs,Girotti:2001ku,Matsuda:1996vq}. This acquired mass, however, is due to $\rho$, while in the usual bosonic model the spontaneous generation of mass occurs due to $\sigma$ acquiring a non-vanishing vacuum expectation value. Therefore, we may say that we do not have anything that we can interpret as a bosonic limit of the spontaneous generation of mass from the SUSY model (since $\rho$ is not present in the usual bosonic model).

The aim of the present paper is to study the phase structure in a deformed nonlinear sigma model. In particular, we are interested in two generalized versions of NLSM, with manifest and softly broken supersymmetry, such that the two situations differ by one single parameter (denoted by $\eta$), which will allow us to have a clearer view of the role of supersymmetry in the dynamical generation of mass. In the next section, we start with the two-dimensional SUSY NLSM, with a more general constraint satisfied by the superfields and proceed to discuss the dynamical generation of mass in this model.

\section{Soft Broken Supersymmetry in (1+1) dimensional SUSY NLSM}

We start with a slight deformation of the SUSY NLSM, introducing a more general constraint for the superfields $\Phi_a$:
\begin{eqnarray}\label{eq1aa}
S=\int\!{d^4z}~\Big{\{}\frac{1}{2}\Phi_a(z) D^2\Phi_a(z) +\Sigma(z)\left[\Phi_a(z)^2-\frac{N}{g} H(z)\right]\Big{\}}~,
\end{eqnarray}

\noindent
where $\Sigma(z)$ is a Lagrange multiplier for the modified constraint $\Phi_a^2(z)=\dfrac{N}{g}H(z)$, where $H(z)$ is a constant superfield which possesses the $\theta$-expansion $H(z)=1-\theta^2~g~\eta$. Note that $H(z)$ breaks SUSY explicitly and we recover the supersymmetric action for the NLSM, Eq. (\ref{eq1}), for $\eta=0$.

The new constraints to the components of the fundamental superfields $\Phi_a$ are:
\begin{eqnarray}\label{constraints2}
\phi_a^2 = \frac{N}{g}~,\hspace{1cm}\psi^\alpha_a\phi_a = 0~,\hspace{1cm} F_a\phi_a = \frac{1}{2}\psi^\beta_a\psi_{a\beta}+g~\eta~.
\end{eqnarray}

In order to study the phase structure of the model, let us start assuming that the N-th component $\Phi_N(x,\theta)$ and $\Sigma$ both have constant non-trivial VEVs given by
\begin{eqnarray}\label{eq2}
\langle\Sigma\rangle&=&\Sigma_{cl}=\rho_{cl}-\theta^2\sigma_{cl}~,\nonumber\\
\langle\Phi_N\rangle&=&\sqrt{N}~\Phi_{cl}=\sqrt{N}~(\phi_{cl}-\theta^2F_{cl})~.
\end{eqnarray}

\noindent
Let us also make a shift in these superfields by redefining $\Sigma\rightarrow(\Sigma+\Sigma_{cl})$ and $\Phi_N\rightarrow \sqrt{N}(\Phi_N+\Phi_{cl})$, so that we can rewrite the action (\ref{eq1aa}) in terms of the new fields as
\begin{eqnarray}\label{eq3}
S&=&\int\!{d^4z}\Big{\{}\frac{1}{2}\Phi_a (D^2+2\Sigma_{cl})\Phi_a 
+\Sigma\left(\Phi_a^2+N\Phi_{cl}^2+2{N}\Phi_{cl}\Phi_N-\frac{N}{g} H(z)\right)\nonumber\\
&&+{N}\Phi_N\left(D^2\Phi_{cl}+2\Phi_{cl}\Sigma_{cl}\right)
+\frac{N}{2}\Phi_{cl} D^2\Phi_{cl}+N\Sigma_{cl}\left(\Phi_{cl}^2-\frac{1}{g}\right)\Big{\}}~.
\end{eqnarray}

We can immediately see that $\Sigma_{cl}$ (i.e., the VEV of the superfield $\Sigma$) gives mass to the fundamental superfields $\Phi_a$, and that this ``mass'' is $\theta$-dependent, therefore generating different masses to the bosonic and fermionic components of $\Phi_a$, showing a possible solution where supersymmetry is broken.

At the leading order, the propagator of $\Phi_a$ must satisfy
\begin{eqnarray}\label{eq4}
[D^2(z)+2\Sigma_{cl}]\Delta(z-z^{\prime})=i\delta^{(4)}(z-z^{\prime})~,
\end{eqnarray}

\noindent
where $\delta^{(4)}(z-z^{\prime})\equiv \delta^{(2)}(x-x^{\prime})\delta^{(2)}(\theta-\theta^{\prime})$, and $\delta^{(2)}(\theta)=-\theta^2$. 

By solving (\ref{eq4}) using the methods described in \cite{Boldo:1999nd,Gallegos:2011ag}, we get the propagator for the superfield $\Phi_a$:
\begin{eqnarray}\label{eq8b}
\Delta(k)&=&-\frac{i}{k^2+4\rho_{cl}^2-2\sigma_{cl}}\Big{\{}
D^2-2\rho_{cl}+\frac{2\sigma_{cl}}{k^2+4\rho_{cl}^2}\Big[
(4\rho_{cl}^2-k^2)\theta^2\nonumber\\
&&+2\rho_{cl}\theta^{\alpha}D_{\alpha}+4\rho_{cl}\theta^2D^2+k^{\alpha\beta}\theta_\alpha D_\beta \Big]\Big{\}}\delta^{(2)}(\theta-\theta^{\prime})~,
\end{eqnarray}
\noindent
which reduces to the usual propagator of a massive scalar superfield for $\sigma_{cl}=0$. 

From Eq.(\ref{eq3}) we can see that there exists a mixing between $\Phi_N$ and $\Sigma$, but this mixing only contributes to the next-to-leading order in the $1/N$ expansion. For now, we can neglect this mixing, since we will deal with the SUSY NLSM only at leading order in $1/N$.

With the propagator of $\Phi_a$, we can evaluate the effective potential through the tadpole method~\cite{Weinberg:1973ua,Miller:1983fe,Miller:1983ri}. At leading order, the tadpole equation for the superfield $\Phi_N$ can be cast as
\begin{eqnarray}\label{eq9}
\frac{\partial U_{eff}}{\partial {\Phi}_{cl}}
=N\left[D^2\phi_{cl}+2\Phi_{cl}\Sigma_{cl}\right]=N\left[F_{cl}+2\phi_{cl}\rho_{cl}-2\theta^2(\phi_{cl}\sigma_{cl}+F_{cl}\rho_{cl})\right]=0,
\end{eqnarray}
\noindent where $U_{eff}$ is the superfield effective superpotential.

\begin{figure}[ht]
\includegraphics[height=4cm ,angle=0 ,width=8cm]{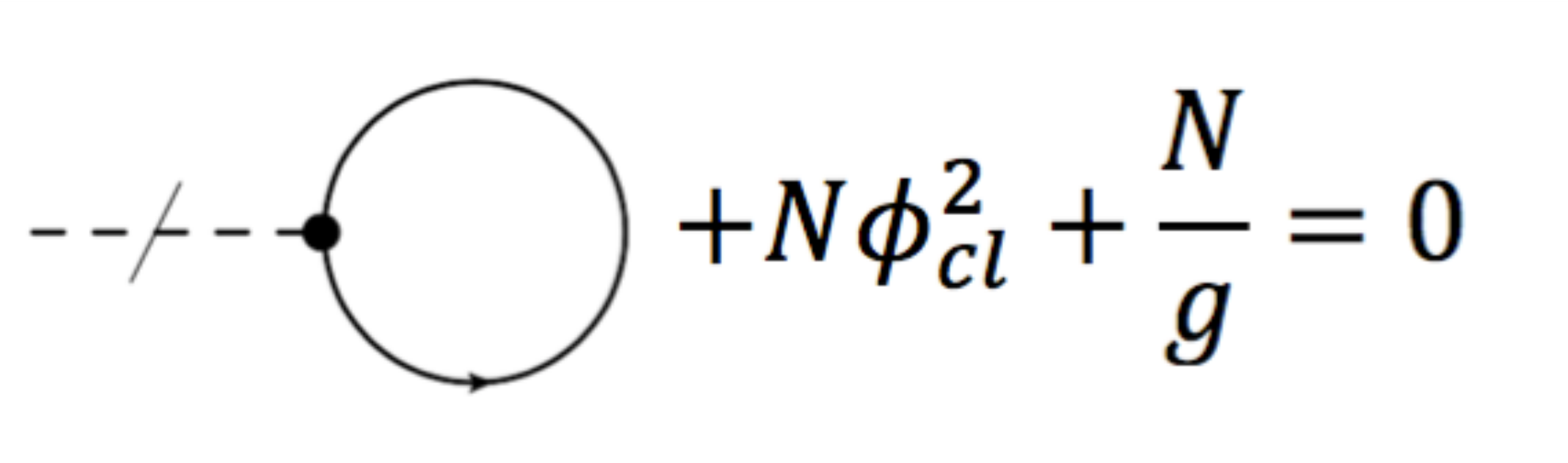}
\caption{Tadpole equation of $\Sigma$ at leading order. Continuous lines represent a $\Phi_a$ propagator and the cut dashed line an external $\Sigma$ propagator.}\label{gap}
\end{figure}

On the other hand, the tadpole equation for $\Sigma$ is (cf. Fig. \ref{gap}): 
\begin{equation}
\frac{\partial U_{eff}}{\partial \Sigma_{cl}}=N\Phi_{cl}^2-\frac{N}{g}H(z) +N\int\!\frac{d^2k}{(2\pi)^2}\Delta(k) = 0.
\end{equation}
Substituting the expression for $\Delta(k)$, and using the fact that $D^2\delta^{(2)}(\theta-\theta)=1$ and $\delta^{(2)}(\theta-\theta)=0$, we obtain
\begin{eqnarray}\label{eq11}
\frac{\partial U_{eff}}{\partial \Sigma_{cl}}&=& N\Phi_{cl}^2-\frac{N}{g}H(z) - iN\int\!\frac{d^2k}{(2\pi)^2}\Big{\{}
\frac{1}{k^2+(4\rho_{cl}^2-2\sigma_{cl})}+\frac{8\sigma_{cl}\rho_{cl}~\theta^2}{[k^2+(4\rho_{cl}^2-2\sigma_{cl})](k^2+4\rho_{cl}^2)}\Big{\}}\nonumber\\
&=&N\Big\{\phi_{cl}^2-\lambda-\frac{1}{4\pi}\ln\left(\frac{4\rho^2_{cl}-2\sigma_{cl}}{\mu^2}\right)
-\theta^2\left[2\phi_{cl}F_{cl}+\frac{\rho_{cl}}{\pi}\ln\left(\frac{4\rho_{cl}^2-2\sigma_{cl}}{4\rho^2}\right)+\eta\right]\Big\}=0,
\end{eqnarray}

\noindent where $\lambda=\dfrac{1}{g}$ is the renormalized coupling and $\mu$ is a mass scale introduced by the regularization by dimensional reduction (i.e., $\int d^2k/(2\pi)^2 \rightarrow \mu^{\epsilon}\int d^{2-\epsilon}k/(2\pi)^{2-\epsilon}$). 

In the tadpole equations, each term of the $\theta$ expansion has to vanish independently, i.e., the classical fields have to satisfy
\begin{eqnarray}
&&F_{cl}+2\phi_{cl}\rho_{cl}=0~,\label{eq12a1}\\ 
&&F_{cl}\rho_{cl}+\phi_{cl}\sigma_{cl}=0~,\label{eq12a2}\\
&&\phi_{cl}^2-\lambda-\frac{1}{4\pi}\ln\left(\frac{4\rho_{cl}^2-2\sigma_{cl}}{\mu^2}\right)=0~,\label{eq12a3}\\
&&2\phi_{cl}F_{cl}+\eta+\frac{\rho_{cl}}{\pi}\ln\left(\frac{4\rho_{cl}^2-2\sigma_{cl}}{4\rho_{cl}^2}\right)=0~.\label{eq12a4}
\end{eqnarray}

With the tadpole equations in hands, the effective potential $V_{eff}=\int{d^2\theta}U_{eff}$ is obtained by integrating Eq.(\ref{eq9}) over $\Phi_{cl}$ and Eq.(\ref{eq11}) over $\Sigma_{cl}$ as 
\begin{eqnarray}\label{eq12a}
-\frac{V_{eff}}{N}&=&\int{d^2\theta}\Big{\{}\int{d\Phi_{cl}}\left[F_{cl}+2\phi_{cl}\rho_{cl}-2\theta^2(\phi_{cl}\sigma_{cl}+F_{cl}\rho_{cl})\right]\nonumber\\
&-&\int{d\Sigma_{cl}}
\Big{[}\phi_{cl}^2-\lambda-\frac{1}{4\pi}\ln\left(\frac{4\rho^2_{cl}-2\sigma_{cl}}{\mu^2}\right)
-\theta^2\left[2\phi_{cl}F_{cl}+\frac{\rho_{cl}}{\pi}\ln\left(\frac{4\rho_{cl}^2-2\sigma_{cl}}{4\rho^2}\right)+\eta\right]
\Big{]} \Big{\}}\nonumber\\
&=&\frac{F_{cl}^2}{2}+\eta \rho_{cl}+\sigma_{cl}\left(\phi_{c}^2+\frac{1}{4\pi}-\lambda\right) +2 F_{cl} \rho_{cl} \phi_{cl}-\frac{\rho_{cl}^2}{2 \pi } \ln\left(\frac{4 \rho_{cl}^2}{\mu^2} \right)\nonumber\\
&&+\frac{(2\rho_{cl}^2-\sigma_{cl})}{4 \pi}\ln\left(\frac{4 \rho_{cl}^2-2 \sigma_{cl} }{\mu ^2}\right)~.
\end{eqnarray}

As we did for the classical action, we can eliminate the auxiliary field $F_{cl}$ using its equation of motion, 
\begin{eqnarray}\label{eomF}
F_{cl}=-2\rho_{cl}\phi_{cl},
\end{eqnarray}

\noindent allowing us to write the effective potential as 
\begin{eqnarray}\label{eq12ccc}
-\frac{V_{eff}}{N}&=&
\eta \rho_{cl}+\sigma_{cl}\left(\phi_{c}^2+\frac{1}{4\pi}-\lambda\right) -2 \rho_{cl}^2 \phi_{cl}^2-\frac{\rho_{cl}^2}{2 \pi } \ln\left(\frac{4 \rho_{cl}^2}{\mu^2} \right)\nonumber\\
&&+\frac{(2\rho_{cl}^2-\sigma_{cl})}{4 \pi}\ln\left(\frac{4 \rho_{cl}^2-2 \sigma_{cl} }{\mu ^2}\right).
\end{eqnarray}

Since $\sigma_{cl}$ is an auxiliary field we may use its equation ($\frac{\partial V_{eff}}{\partial \sigma_{cl}}=0$) to find $\sigma_{cl} = 2\rho_{cl}^2-\frac{\mu^2}{2}\mathrm{e}^{-4\pi\lambda}$ and write $V_{eff} (\phi_{cl}, \rho_{cl})$, from which we derive the conditions that extremize the effective potential:
\begin{eqnarray}
&&\frac{\partial V_{eff}}{\partial \phi_{cl}}=\mu^2 \phi_{cl} \mathrm{e}^{4\pi (\phi_{cl}^2 - \lambda)} = 0~,\label{gap1}\\
&&\frac{\partial V_{eff}}{\partial \rho_{cl}}=\eta - 4\lambda\rho_{cl} - \frac{\rho_{cl}}{\pi} \ln \left(\frac{4\rho_{cl}^2}{\mu^2}\right)= 0.\label{gap2}
\end{eqnarray}
Solving these equations, we determine two critical points:
\begin{eqnarray} 
&&\phi_{cl}=0, \hspace{.3cm} \rho_{cl}^\pm=\frac{\mu \mathrm{e}^{-2\pi\lambda}}{2}\frac{\beta}{W(\pm \beta)} = (\pm) \frac{\mu}{2} \mathrm{e}^{-2\pi\lambda + W(\beta)},\hspace{.3cm}\sigma_{cl}=\frac{\mu^2}{2}\mathrm{e}^{-4\pi\lambda}\left(\mathrm{e}^{2W(\pm \beta)}-1\right)~\label{eq15b}
\end{eqnarray}
where $\beta = \frac{\eta\pi}{\mu} \mathrm{e}^{2\pi\lambda}$ and $W(\beta)$ is the Lambert's $W$-function, and we have used  $W(\pm \beta) = \pm \beta \mathrm{e}^{- W(\beta)}$.

In order to determine if those critical points are minima, we compute the Hessian of $V_{eff} (\phi_{cl}, \rho_{cl})$:
\begin{equation}
H = \left(\begin{array}{cc}
\mu^2 N \mathrm{e}^{-4\pi(\phi_{cl}^2 - \lambda)} (1+8\pi \phi_{cl}^2) & 0 \\
0 & \frac{2N}{\pi}\left(1+2\pi\lambda + \ln\left(\frac{4\rho_{cl}^2}{\mu^2}\right)\right).
\end{array}\right)
\end{equation}
At the critical point, $\phi_{cl}=0$, so we have:
\[
\det H = \frac{2N\mu^2}{\pi}\mathrm{e}^{-4\pi\lambda}\left(1+2\pi\lambda + \ln\left(\frac{4\rho_{cl}^2}{\mu^2}\right)\right).
\]

\noindent We can see that the condition $\det H>0$ is satisfied for $\left[1+2\pi\lambda + \ln\left(\frac{4\rho_{cl}^2}{\mu^2}\right)\right]>0$.

Such solution is $O(N)$ symmetric, presenting a dynamical generation of mass to the fundamental matter fields $\phi$ and $\psi$, which are given by 
\begin{eqnarray}\label{eq1da}
M^2_{\phi}=4\langle\rho\rangle^2-2\langle\sigma\rangle=\mu^2\mathrm{e}^{-4\pi\lambda}~,\hspace{1cm} M^2_{\psi} = 4\langle\rho\rangle^2=\mathrm{e}^{2W(\pm \beta)}M^2_{\phi}~.
\end{eqnarray}

\begin{figure}[ht]
\includegraphics[height=8cm ,angle=0 ,width=12cm]{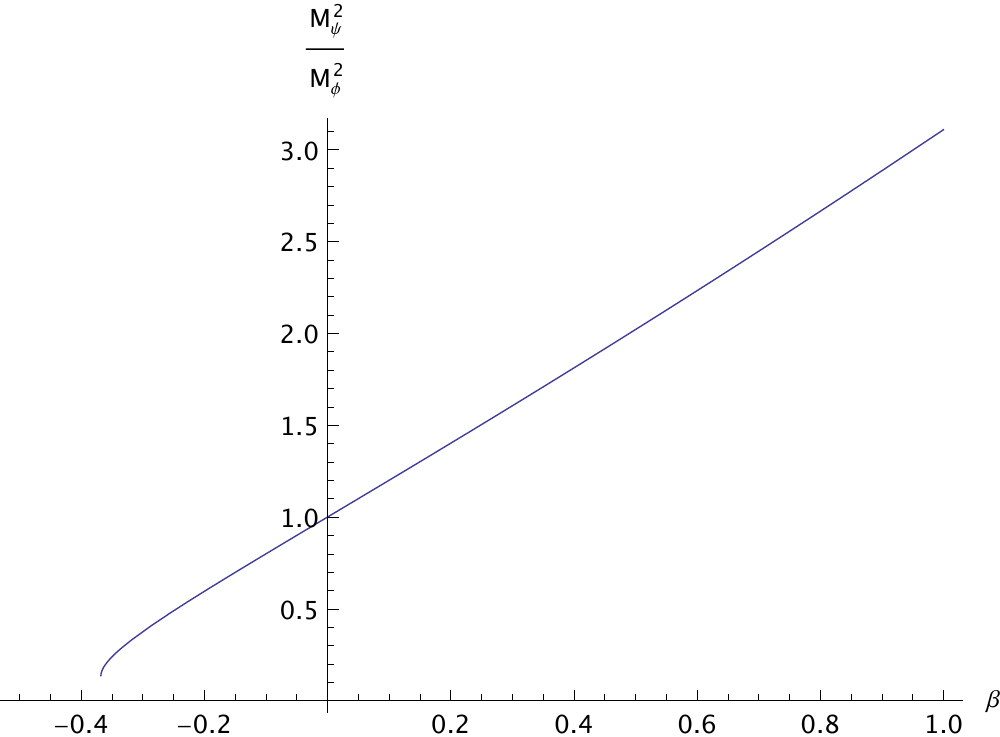}
\caption{Mass rate $M^2_{\psi}/M^2_{\phi}$ as function of $\beta=\frac{\eta\pi}{\mu} \mathrm{e}^{2\pi\lambda}$.}\label{massrate}
\end{figure}

The Lambert's $W$-function assumes its lowest real value for $\beta=-1/\mathrm{e}$, where the mass rate becomes $M^2_{\psi}/M^2_{\phi}=\mathrm{e}^{-2}$. In the Figure \ref{massrate} we plot the mass rate $M^2_{\phi}/M^2_{\psi}$ as function of $\beta$. It is easy to see that if we take the limit $\eta\rightarrow 0$ ($\beta\rightarrow0$), we recover the supersymmetric solution 
\begin{eqnarray} 
&&\phi_{cl}=0, \hspace{.3cm} \rho_{cl}=\pm\frac{\mu^2}{2}\mathrm{e}^{-2\pi\lambda}~,\hspace{.3cm}
\sigma_{cl}=0,\label{eq15bb}\end{eqnarray}
\noindent with $M^2_{\phi}=M^2_{\psi}$.

\section{Final remarks}

Summarizing, we study the dynamical generation of mass in a deformed $D=(1+1)$ supersymmetric nonlinear sigma model, where the deformation is introduced by imposing a soft supersymmetry breaking constraint. We showed that the generated masses to the matter fields $\phi$ and $\psi$ are $M^2_{\phi}=\mu^2\mathrm{e}^{-4\pi\lambda}$ and $M^2_{\psi} =\mathrm{e}^{2W(\pm \beta)}M^2_{\phi}$, respectively. We see that supersymmetry is broken for a nonvanishing deformation parameter $\eta$, while $O(N)$ symmetry is kept manifest. Supersymmetry is restored in the limit $\eta\rightarrow0$ (that is, $\beta\rightarrow0$). 

As we have mentioned, the usual NLSM \eqref{eq0a} can be obtained from \eqref{eq1} through the so-called bosonic limit, by setting $\psi=\rho=\chi=0$, and $\sigma \neq 0$ in \eqref{eq1a}. However, such soft transition to the bosonic model is not observed for the mass acquired by the fields in the symmetric phase. In fact, in the ordinary bosonic case the spontaneous generation of mass occurs due to $\sigma$ acquiring a non-vanishing vacuum expectation value, while in the manifest supersymmetric solution, the generated mass is due to $\rho$, a field not present in the bosonic NLSM. In that sense, we may say that we do not have anything that we can interpret as a bosonic limit of the spontaneous generation of mass from the supersymmetric model. 

Our work has explored a deformed model given by action \eqref{eq1aa}, where the supersymmetry is softly broken. We found that even in that deformed model such bosonic limit of the spontaneous generation of mass is absent. However, unlike the supersymmetric model, where the fermionic acquired mass is independent of the VEV of $\sigma$, we found that in the deformed model with softly broken SUSY the generated mass of fermionic fields depend on a nonvanishing $\langle\sigma\rangle$.

Finally, we expect that gauge and noncommutative (with constant noncommutativity parameter) extensions of this model, such as Noncommutative SUSY CP$^{(N-1)}$~\cite{Ferrari:2006xx}, should exhibit same properties of the present model, since the tadpole diagrams in noncommutative theories are the same of the commutative ones. 

\vspace{.5cm}
{\bf Acknowledgments.} This work was partially supported by the Brazilian agency Conselho Nacional de Desenvolvimento Cient\'{\i}fico e Tecnol\'{o}gico (CNPq). A.C.L. has been partially supported by the CNPq project 307723/2016-0 and 402096/2016-9. AJS has been partially supported by the CNPq project 306926/2017-2.

\newpage 

\end{document}